\documentclass[plb,preprint,tightenlines,floatfix,showpacs,preprintnumbers,nofootinbib]{revtex4}
 \usepackage[dvips,final]{graphicx}
  \usepackage{amssymb}
   \usepackage{amsmath}
    \usepackage{epsfig}
     \usepackage{bm}% bold math
      \usepackage{pifont}
\usepackage{color}

\def\agoth{\relax\ifmmode{\mathfrak A}\else{${\mathfrak A}${ }}\fi}
\def\muF{\relax\ifmmode\mu_\text{F}^2\else{$\mu_\text{F}^2${ }}\fi}
\def\muR{\relax\ifmmode\mu_\text{R}^2\else{$\mu_\text{R}^2${ }}\fi}
\def\muO{\relax\ifmmode{\mu_{0}^{2}}\else{$\mu_{0}^{2}${ }}\fi}
\def\Mev{\relax\ifmmode{\text{MeV}}\else{MeV{ }}\fi}

\def\Li{\relax\ifmmode{\textbf{Li}_{2}}\else{Li$_2${ }}\fi}
\def\Im{\relax{\textbf{Im}{}}}

\newcommand{\gev}[1]{\relax\ifmmode{\text{GeV}^{#1}}\else{GeV$^{#1}${ }}\fi}

\def\asb{\relax\ifmmode \bar{\alpha}_s\else{$ \bar{\alpha}_s${ }}\fi}
\def\as{\relax\ifmmode \alpha_s\else{$ \alpha_s${ }}\fi}
\def\acal{\relax\ifmmode{\cal A}\else{${\cal A}${ }}\fi}

\def\as{\relax\ifmmode \alpha_s\else{$ \alpha_s${ }}\fi}  %%%%%%%%%
\def\abar{\relax\ifmmode{\bar{a}}\else{$\bar{a}${ }}\fi}  %%%%%%%%%
\newcommand{\be}{\begin{equation}}
\newcommand{\ee}{\end{equation}}

\newcommand{\ba}{\begin{eqnarray}}
\newcommand{\ea}{\end{eqnarray}}
\newcommand{\bg}{\begin{gather}}
\newcommand{\foma}{\end{gather}}

\newcommand{\vecc}[1]{\mbox{\boldmath $#1$}}

\def\halb{\frac{1}{2}}
\def\e{\epsilon}

\def\w{\omega}

\def\pd{\partial}

\def\F{\Phi}

\def\pb{\bar \psi}

\def\ex{\hbox{e}}
\def\S{\Sigma}
\def\F{\Phi}
\def\<{\langle}
\def\>{\rangle}

\def\a{\alpha}

\def\g{\gamma}  \def\G{\Gamma}
\def\d{\delta}  
   
\def\s{\sigma}
  
\def\x{\xi}

\def\m{\mu}
\def\n{\nu}
\def\t{\tau}

\def\w{\omega}

\def\({\left(}
\def\[{\left[}
\def\){\right)}
\def\]{\right]}

\def\pd{\partial}

\def\pa{{\cal P}}

\begin{document}
\thispagestyle{empty}
\date{\today}
\preprint{\hbox{RUB-TPII-04/09}}
%\vspace*{-10mm}

\title{Renormalization-group properties of transverse-momentum
       dependent parton distribution functions in the light-cone
        gauge with the Mandelstam-Leibbrandt prescription}
\author{I.~O.~Cherednikov}
\email{igor.cherednikov@jinr.ru}
\affiliation{Bogoliubov Laboratory of Theoretical Physics,
             Joint Institute for Nuclear Research, \\
             RU-141980 Dubna, Russia}
\affiliation{INFN Gruppo collegato di Cosenza, \\
             Dipartimento di Fisica, Universit$\grave{a}$
             della Calabria, \\ I-87036 Arcavacata di Rende, Italy}
\affiliation{Institute for Theoretical Problems of Microphysics,
             Moscow State University, \\ RU-119899, Moscow, Russia}
\author{N.~G.~Stefanis}
\email{stefanis@tp2.ruhr-uni-bochum.de}
\affiliation{Institut f\"{u}r Theoretische Physik II,
             Ruhr-Universit\"{a}t Bochum, \\
             D-44780 Bochum, Germany}
\affiliation{Bogoliubov Laboratory of Theoretical Physics,
             Joint Institute for Nuclear Research, \\
             RU-141980 Dubna, Russia}
\vspace {10mm}
%\cleardoublepage
\begin{abstract}
The renormalization-group properties of transverse-momentum
dependent parton distribution functions in the light-cone gauge
with the Mandelstam-Leibbrandt prescription for the gluon
propagator are addressed.
An expression for the transverse component of the gauge field
at light-cone infinity, which plays a crucial role in the
description of the final-/initial-state interactions
in the light-cone axial gauge, is obtained.
The leading-order anomalous dimension is calculated in this gauge
and the relation to the results obtained in other gauges is worked
out.
It is shown that, using the Mandelstam-Leibbrandt prescription,
the ensuing anomalous dimension does not receive contributions from
extra rapidity divergences related to a cusped junction point of
the Wilson lines.
\end{abstract}
%Keywords: Gauge invariance
%          TMD parton distribution functions
%          Wilson lines
%          Phase factors
%          Single spin asymmetries
%          Renormalization and anomalous dimensions
\maketitle

%\tableofcontents

%%%%%%%%%%%%%%%%%%%%%%%%%%%%%%%%%%%%%%%%%%%%%%%%%%%%%%%%%%%%%%%%%%%%%%%
%%%%%%%%%%%%%%%%%%%%%%%%%%%%%%%%%%%%%%%%%%%%%%%%%%%%%%%%%%%%%%%%%%%%%%%
%\cleardoublepage

\section{Introduction}
\label{sec:intro}

Parton distribution functions (PDF)s contain nonperturbative
information about the intrinsic structure of hadrons in terms of their
 constituents---quarks and gluons \cite{Col03,BR05,Col08}.
In completely inclusive processes (e.g., deeply inelastic scattering
(DIS), where the hard virtual photon with momentum $q^2 = - Q^2$
probes a hadron $h$ with momentum $P$), integrated PDFs $f_i (x, Q^2)$
($i$ marking the sort of parton)
depend on the longitudinal-momentum fraction $x$, which becomes equal
to the Bjorken variable $x_{\rm B}$ in the limit
$Q^2 \to \infty$, $x_{\rm B} = Q^2/2(q\cdot P) = \hbox{const}$,
and on the scale of the hard subprocess $Q^2$.
In the Bjorken limit, these distributions (Feynman parton densities)
are related to the (unpolarized) quark and antiquark structure
functions
\begin{equation}
  F_1 (x_{\rm B}, Q^2)
  =
  \frac{1}{2x_{\rm B}} \ F_2 \(x_{\rm B}, Q^2\)
=
  \frac{1}{2} \sum_i \ e_i^2
  \left[
        f_i(x_{\rm B}, Q^2) + {\bar f}_{i}(x_{\rm B}, Q^2)
  \right]
\label{eq:iPDF_def}
\end{equation}
%Eq (1)
in the leading-twist approximation and in leading order of the
coupling $\alpha_s$ (where $e_i$ is the electric charge of the
quark of flavor $i$) \cite{AP77}.
Equation (\ref{eq:iPDF_def}) originates from the DIS factorization
expression
\begin{equation}
  F_1 (x_{\rm B}, Q^2)
=
  \sum_i \ \int_{x_{\rm B}}^1\! \frac{dz}{z} \
  C_{1i} \[\frac{x_{\rm B}}{z}, \frac{Q^2}{\mu^2}, \alpha_s (\mu) \] \
  \[f_i(z, \mu^2) + {\bar f}_{i}(z, \mu^2) \] \ ,
\label{eq:iPDF_coeff}
\end{equation}
%Eq (2)
where the perturbative coefficient functions $C_{1i}$ are taken in
leading order (LO):
$C_{1i}^{\rm LO} = \frac{1}{2} e_i^2 \ \delta(x/z -1)$.
It can be considered as a triumph of perturbative QCD that the QCD
evolution correctly describes the logarithmic dependence of the parton
densities on the hard scale $Q^2$ and is, therefore, able to explain
the experimentally observed violation of the Bjorken scaling.
Moreover, the gauge-invariant operator definition of integrated
PDFs
\begin{eqnarray}
  f_{i/h} (x, Q^2)
=
  \frac{1}{2} \int \frac{d\x^-}{2\pi}
  \ex^{- i k^+ \x^-} { \< h | } \pb_i
  (\x^- ) {\cal P} \exp
  \left[-ig\int_{0^-}^{\xi^-}\!\! dz^{\mu} A^{a}_{\mu}(z) t^{a}
  \right] \g^+
  \psi_i(0^-) { |h\> }\Bigg|_{\xi^+,\, \xi_\perp = 0}
\label{eq:ipdf}
\end{eqnarray}
%Eq (3)
with $k^+ = x P^+$ allows one to relate their moments to the matrix
elements of the twist-two operators arising in the operator product
expansion (OPE) on the light-cone \cite{CS82}.
The renormalization properties of these PDFs are governed by the
DGLAP equation \cite{AP77,DGLAP}, establishing the logarithmic
dependence on $Q^2$ mentioned above.

The study of semi-inclusive processes, such as semi-inclusive deeply
inelastic scattering (SIDIS), or the Drell-Yan
(DY) process---where one more final or initial hadron is detected and
its transverse momentum is observed---requires the introduction of
more complicated quantities, viz., unintegrated, i.e.,
transverse-momentum dependent (TMD) distribution or fragmentation
functions:
\begin{equation}
  \hbox{semi-inclusive} \ \
\to \ \
  f_{i/h} (x, \vecc k_\perp, Q^2) \ .
\end{equation}
%Eq (4)
The most natural generalization of the operator definition
(\ref{eq:ipdf}), respecting gauge invariance and collinear
factorization, reads \cite{CS81,CS82,Col03,JY02,BJY03,BMP03}
\begin{eqnarray}
  && f_{i/h} \left(x, \mbox{\boldmath$k_\perp$};\mu^2 \right)
=
  \frac{1}{2}
  \int \frac{d\xi^- d^2\mbox{\boldmath$\xi_\perp$}}{2\pi (2\pi)^2}
  {\rm e}^{-ik^{+}\xi^{-} +i \mbox{\footnotesize\boldmath$k_\perp$}
\cdot \mbox{\footnotesize\boldmath$\xi_\perp$}}
  \left\langle
              h |\bar \psi_i (\xi^-, \mbox{\boldmath$\xi_\perp$})
              [\xi^-, \mbox{\boldmath$\xi_\perp$};
   \infty^-, \mbox{\boldmath$\xi_\perp$}]^\dagger
\right. \nonumber \\
&& \left. \times
   [\infty^-, \mbox{\boldmath$\xi_\perp$};
   \infty^-, \mbox{\boldmath$\infty_\perp$}]^\dagger
   \gamma^+[\infty^-, \mbox{\boldmath$\infty_\perp$};
   \infty^-, \mbox{\boldmath$0_\perp$}]
   [\infty^-, \mbox{\boldmath$0_\perp$}; 0^-,\mbox{\boldmath$0_\perp$}]
   \psi_i (0^-,\mbox{\boldmath$0_\perp$}) | h
   \right\rangle \ \, ,
\label{eq:tmd_naive}
\end{eqnarray}
%Eq (5)
where gauge invariance is ensured by means of the path-ordered
Wilson-line operator (gauge link) with the generic form
\begin{equation}
  { [y,x|\Gamma] }
=
  {\cal P} \exp
  \left[-ig\int_{x[\Gamma]}^{y}dz_{\mu} A_{a}^{\mu}(z) t_{a}
  \right] \ .
\label{eq:link}
\end{equation}
%Eq (6)
The transverse gauge links extending to light-cone infinity are also
included in (\ref{eq:tmd_naive}) and the dependence on $Q^2$ is
taken into account via the renormalization-group (RG).

However, as it has been pointed out in \cite{CS81}, when one retains
in the parton densities the intrinsic transverse momentum, extra
undesirable divergences appear.
These divergences are associated with the particular features of the
light-cone gauge (or the use of purely lightlike Wilson lines) that
must be removed by some consistent method (see, e.g.,
\cite{Col08,CRS07,Bacch08,CS07,CS08}).
For instance, they can be avoided by using non-lightlike gauge links in
covariant gauges, or by employing an axial gauge, but going
off-the-light-cone \cite{CS81,JMY04}.
This involves the introduction of an extra rapidity parameter and
entails an additional evolution equation \cite{CS81}, rendering the
reduction to the integrated PDF questionable.

Another strategy, based on a subtraction formalism of these extra
divergences in terms of a ``soft'' factor (defined as the vacuum
average of particular Wilson lines and amounting to a generalized
renormalization of TMD PDFs), was presented in Refs.\
\cite{CH00,Hau07,CM04,CS07,CS08}.
The major finding in our previous investigations in \cite{CS07,CS08}
was that, adopting the light-cone gauge, the leading gluon radiative
corrections associated with the transverse gauge link were found to
give rise to an extra term in the anomalous dimension of the TMD PDF
that exhibits a $\ln p^+$ behavior---characteristic of a contour
with a cusp.
The new definition for the TMD PDF, proposed in these works, has two
important advantages: (i) it reduces to the correct integrated case
and (ii) it coincides with the result obtained in the Feynman
gauge that is untainted by contour obstructions.

In (non-covariant) axial gauges the partonic interpretation of the
distribution functions is preserved because the gauge links can be set
to unity under the gauge condition.
It is well-known that among the axial light-cone gauges there is one
which has important advantages.
Indeed, employing the light-cone gauge in association with the
Mandelstam-Leibbrandt (ML) prescription
\cite{Man83,Lei84}\footnote{Let us call in what follows the
light-cone gauge with this prescription the ``ML-gauge''.} in the
calculation of the quark self-energy and the quark-quark-gluon vertex
\cite{LN83, BA96}, and also the DGLAP kernel in NLO \cite{BHKV98}, it
was shown that no undesirable singularities appear---even at the
intermediate steps of the calculation---in contrast to the
principal-value (PV) prescription used in \cite{CFP80}.
In the ML-gauge, the contributions of the real and the virtual
diagrams are well-defined in the ``end-point'' region separately.
[This is the region proportional to the delta-function of the
longitudinal fraction of the hadron's momentum $\sim \delta(1-x)$].
In the calculation of the evolution of the inclusive (integrated) PDF,
this is not important---at least in LO---since all these singularities
cancel in the final result.
But the situation changes for the unintegrated TMD PDFs.
In that case, the spacelike distance between the quark operators in the
corresponding matrix elements acts like an ultraviolet (UV)-regulator,
thus preventing the mutual cancelation of the extra
(mixed)\footnote{We distinguish between ``pure'' singular terms, having
only a single UV or rapidity divergence, and ``mixed'' ones, which
contain two poles of different origin simultaneously.}
divergent terms between the virtual and the real gluon contributions.
As a result, the rapidity divergences contributing to the UV-divergent
part of the self-energy graph remain uncanceled.
Exactly those terms of the splitting function, arising from the
``end-point'' region, give rise to extra (mixed) UV-singularities in
the TMD function that can be eliminated by employing the
ML-prescription---even extending the calculation to the NLO
\cite{BHKV98}.

\begin{figure}
\centering
\includegraphics[scale=0.7,angle=90]{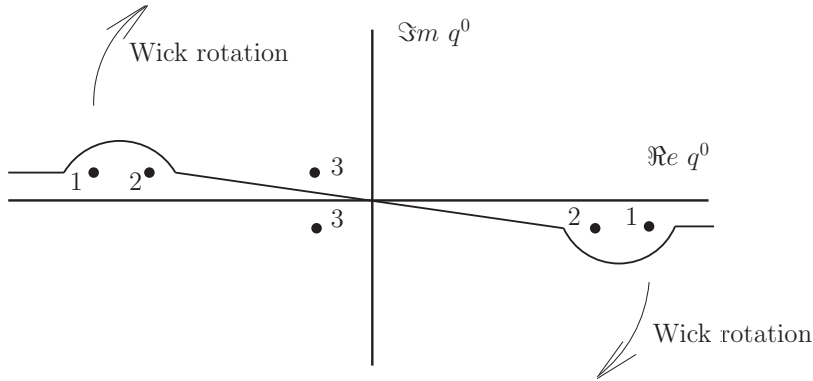}~~
\caption{Integration contour and poles in the
         $( \hbox{\bf Re} \ q^0, \Im \ q^0)$
         plane: the poles of the gluon propagator using the
         ML-prescription (position 1) and those in a covariant gauge
         (position 2) belong to the same, i.e., second and fourth,
         quadrants.
         This is in contrast to the poles pertaining to the
         principal-value prescription (position 3).
         The Wick rotation can be performed without changing the
         position of the poles.
\label{fig:ML_poles}}
\end{figure}

From the field-theoretical point of view, the ML-gauge has very
attractive properties as well.
Due to the position of the poles in the same quadrants as for the free
gluon propagator (see Fig. \ref{fig:ML_poles}), one can readily perform
the Wick rotation to the Euclidean space.
This is not possible for the $q^-$-independent prescriptions.
Thus, in the ML-gauge, the standard power counting rules allow one to
estimate the UV-divergences.
Moreover, is has been shown that the ML-prescription arises naturally
in a consistent quantization procedure \cite{BDLS84, BDS87, SF87}.

%%%%%%%%%%%%%%%%%%%%%%%%%%%%%%%FIGURE 1%%%%%%%%%%%%%%%%%%%%%%%%%%%%%%%%
%%%%%%%%%%%%%%%%%%%%%%%%%%%%%%%%%%%%%%%%%%%%%%%%%%%%%%%%%%%%%%%%%%%%%%%

We have pointed out recently \cite{CS07,CS08} that the
renormalization-group properties of TMD PDFs can be profitably analyzed
in terms of their UV anomalous dimensions.
The main reason is that anomalous dimensions are \emph{local}
quantities originating from the geometrical obstructions of the gauge
contours: endpoints, cusps, or self-intersections.
Within this context, gauge-invariant quantities have to fulfil
anomalous-dimension sum rules that represent logarithmic, i.e.,
additive, versions of the Slavnov-Taylor identities
\cite{CS08}.
In our previous works all the calculations have been done in the
light-cone axial gauge with additional $q^-$-independent pole
prescriptions, notably, the advanced, retarded, and the principal-value
one.
It was shown that the extra contribution to the anomalous dimension of
the TMD PDF, given by Eq.\ (\ref{eq:tmd_naive}), can be identified (at
least in the one-loop order) with the well-known cusp anomalous
dimension \cite{KR87}.
In the present investigation we apply this type of approach to a
pole configuration of the gluon propagator controlled by the ML
prescription.
In what follows, we shall first analyze
(Sec.\ \ref{sec:trans-gauge-link-ML}) the behavior of the transverse
component of the (``classical'') gauge field at light-cone
infinity---required for the derivation of the transverse gauge link
that eliminates the residual gauge freedom (after fixing the gauge by
$A^+ = 0$)---and derive an explicit expression for the gauge field in
the ML-gauge.
Then, we shall calculate the UV-divergent parts and the corresponding
anomalous dimension of the TMD PDF (Sec.\ \ref{sec:anom-dim-ML}).
It is remarkable that the result obtained this way is free of
undesirable terms related to contour obstructions and coincides with
the double anomalous dimension of the fermion field.
As we shall show later (Sec.\ \ref{sec:trans-gauge-link}), the crucial
point in verifying the validity of the modified definition of TMD PDFs,
Eq.\ (\ref{eq:tmd_re-definition}) below, proposed in \cite{CS07,CS08},
is that the so-called mixed rapidity divergences are absent in the
ML-gauge, while the contribution of the soft factor (which has been
introduced in order to cancel these divergences in the case of
$q^-$-independent prescriptions) reduces to unity in the one-loop
order---see Sec.\ \ref{sec:soft-factor}.
Our conclusions are presented in Sec.\ \ref{sec:concl}.

\section{Transverse gauge field at light-cone infinity in
         the ML-gauge}
\label{sec:trans-gauge-link-ML}

It was argued in Refs. \cite{JY02,BJY03} that in order to restore the
contribution of the gluon exchanges between the struck quark and the
spectator in the light-cone gauge, one has to take into account the
accumulation of the corresponding phase in the transverse direction
[Note that this phase is suppressed in covariant gauges].
It is precisely this phase that yields the additional transverse gauge
link at light-cone infinity, introduced in Refs.\
\cite{JY02, BJY03, BMP03}, the reason being that this phase is
accumulated very slowly.
Below, we derive an expression for the gauge field in this situation by
adopting the ML-prescription.
This has not been considered before in the literature.

We commence our analysis by calculating the gauge field, the source of
which is a ``classical'' current
\begin{equation}
  j_\m (y)
=
  g \int\! d y'_\m \ \d^{(4)} (y - y')\, , \quad
  y'_\m = v_\m \ \t \ ,
\label{eq:current-2}
\end{equation}
%Eq (7)
corresponding to a charged point-like particle (e.g., a struck quark
in a SIDIS process) and moving with the quasi-constant four-velocity
$v_\m$ along the straight line $v_\m \tau$.
Note that the velocity changes only at the origin, where the sudden
collision with the hard photon takes place and the quark is derailed
to its new ``trajectory''.
The gauge field related to such a current is given by
\begin{equation}
  A^\m (\x)
=
  \int\! d^4 y \ {\cal  D}^{\m\n}(\x - y) j_\n (y)\ \  ,
\label{eq:source1}
\end{equation}
%Eq (8)
where $D^{\m\n}$ is the gluon Green's function.
We assume that the velocity of the struck quark is parallel to the
``plus''- and the ``minus''- light-cone vectors $n^{\pm}_{\mu}$ before
and after the hard collision, respectively:
\begin{eqnarray}
  j_\m (y)
& = &
    g \[ n^+_\m \int_{-\infty}^0 \! d\t \ \d^{(4)}(y - n^+ \t)
  + n^-_\m  \int_{0}^\infty \! d\t \ \d^{(4)}(y - n^+ \t)\]
\nonumber \\
& = &
    g \ \d^{(2)} (\vecc y_\perp) \[ n^+_\m \d(y^-)  \int\!
    \frac{dq^-}{2\pi} \frac{\ex^{-iq^- y^+}}{q^- + i0}
    - n^-_\m \d(y^+)  \int\! \frac{dq^+}{2\pi}
    \frac{\ex^{-iq^+ y^-}}{q^+ - i0}\] \ .
\label{eq:current-3}
\end{eqnarray}
%Eq (9)
Using the results of Refs. \cite{CS07,CS08}, we write
\begin{equation}
  A^\mu (\x)
=
  - g \ n_\n^+ \int\! \frac{d^4 q}{2(2\pi)^4}
   \ {\ex^{- i q \cdot \x}} \ \tilde {\cal D}^{\mu \n} (q)
   \int\! dy^+ dy^- d^2 y_\perp {\ex^{ i q \cdot y}} \d(y^-)
                                \d^{(2)} (\vecc y_\perp)
 \ ,
\label{eq:perp_sour1}
\end{equation}
%Eq (10)
where the free gluon propagator in the light-cone gauge $A^+ = 0$ has
the form
\begin{equation}
  {\cal D}^{\m\n} (z)
=
   \int\! \frac{d^4 q}{(2\pi)^4} \
   {\ex^{- i q \cdot z}} \tilde {\cal D}^{\m\n}(q)
= - \int\! \frac{d^4 q}{(2\pi)^4} \
    \frac{\ex^{- i q \cdot z}} {
    q^2+i0}
    \( g^{\m\n}
  - \frac{q^{\mu}(n^-)^{\nu}+q^{\nu}(n^-)^{\mu}}{[q^+]}
    \)
\label{eq:gluon-prop}
\end{equation}
%Eq (11)
with the square bracket being used in order to remind that $[q^+]$ has
yet to be defined.
Here, and in what follows, we neglect the quark and gluon masses
$m$ and $\lambda$, since we are mainly interested in the
UV-singularities.

Performing the integration over the variable $y$, we get
\begin{equation}
  A^\mu (\x)
=
  - g \ n_\n^+ \int\! \frac{d^4 q}{2(2\pi)^4}
   \ {\ex^{- i q \cdot \x}} \ \frac{\delta(q^-)}{q^2+ i0} \
   \( g^{\m\n}
  - \frac{q^{\mu}(n^-)^{\nu}+q^{\nu}(n^-)^{\mu}}{[q^+]}
    \) \ .
\label{eq:delta}
\end{equation}
%Eq (12)
Before continuing, let us first verify that the longitudinal
(light-cone) components of the gauge field vanish and that there is no
contradiction with the gauge condition.
The ``plus''-component reads
\begin{equation}
  A^+
=
  (A^\mu \cdot n^-_\mu)
  \sim
  n^-_\mu
   \( (n^+)^\mu
  - \frac{q^{\mu} + q^- (n^-)^{\mu}}{[q^+]}
    \)
=
    1 - \frac{q^+}{[q^+]} \ .
\end{equation}
%Eq (13)
Notice that one has for both types of pole prescriptions:
$q^-$-independent, as well as for $q^-$-dependent ones
(like the ML-prescription), $q^+/[q^+] = 1$
(see, e.g.,  Ref. \cite{CDL83}), so that
$$
  A^+
  =
  0 \ ,
$$
a result in agreement with the light-cone gauge.
One the other hand, for the ``minus''-component, we start with
\begin{equation}
A^-
=
  (A^\mu \cdot n^+_\mu)
  \sim
  n^+_\mu
   \( (n^+)^\mu
  - \frac{q^{\mu} + q^- (n^-)^{\mu}}{[q^+]}
    \)
=
    0 - \frac{2 q^-}{[q^+]} \ ,
\end{equation}
%Eq (14)
and carrying out the integration over $q^-$ in Eq.\ ({\ref{eq:delta}})
with $\delta(q^-)$, we find
$$
  A^-
=
  0 \ .
$$

Now turn to the evaluation of the transverse components.
Let us recall that employing $q^-$-independent prescriptions,
the transverse component of the gauge field is
\begin{equation}
   \vecc A^\perp (\infty^-; \vecc \x_\perp) = \frac{g}{4\pi} C_\infty\
   \vecc \nabla^\perp \ln \ \Lambda |\vecc \x_\perp| \, ,
\label{eq:ret}
\end{equation}
%Eq (15)
where the numerical constant $C_\infty$ depends on the pole
prescription according to (see \cite{BJY03})
\begin{equation}
  C_\infty
=
  \left\{
  \begin{array}{ll}
  & \ \ 0  \ , \ {\rm Advanced}    \\
  & - 1 \ , \ {\rm Retarded}  \\
  & - \frac{1}{2} \ , \ {\rm Principal~Value~}\ .
  \
  \end{array} \right.
\label{eq:c_inf}
\end{equation}
%Eq (16)
Here $\Lambda$ is an IR-regulator that does not enter the final
results.

The ML-prescription, being dependent on both variables $q^+$ and $q^-$,
gives rise to a more complicated pole structure in the complex $q^0$
plane, viz.,
\begin{equation}
  \frac{1}{[q^+]_{\rm ML}}
= \left\{
  \begin{array}{ll}
  & \ \ \frac{1}{q^+ + i0q^-}     \\
  & \ \ \frac{q^-}{q^+q^- + i0}
  \
  \end{array} \right.
  \ .
\label{eq:MLdef}
\end{equation}
%Eq (17)
The two possible forms of this prescription, displayed in
Eq.\ (\ref{eq:MLdef}), are, in fact, equivalent to each other.
After performing the integral over $y$ in
Eq.\ (\ref{eq:perp_sour1}), one finds
\begin{equation}
  \vecc A^\perp (\x)
=
  - g \ \pi \int\! \frac{d^4 q}{2(2\pi)^4}
   \ {\ex^{- i q \cdot \x}} \ \frac{\delta(q^-)}{q^2+ i0} \
   \( (n^+)^\perp
  - \frac{q^{\perp} + q^{-}(n^-)^{\perp}}{[q^+]_{\rm ML}}
    \) \ .
\label{eq:delta_perp}
\end{equation}
%Eq (18)
Taking into account that the light-cone vectors $n^\pm$ have only
longitudinal components and separating out the transverse integrations,
we get
\begin{equation}
  \vecc A^\perp (\x)
=
   g \ \pi \int\! \frac{d^2 q_\perp}{(2\pi)^2}
   \ \vecc q^\perp \ \ex^{i q_\perp \xi_\perp}
   \int\! \frac{dq^+dq^-}{(2\pi)^2} \
   \delta(q^-)  \
   \frac{\ex^{- i (q^+ \x^- + q^- \xi^+)}}{( q^2+ i0 )[q^+]_{\rm ML}} \
   \ .
\label{eq:delta_perp_2}
\end{equation}
%Eq (19)

In order to compute the longitudinal part of this expression, we use
for the denominator the $\alpha$-representation and employ for the
delta-function the integral representation
$$
  \delta (q^-) = \frac{1}{2\pi} \int_{-\infty}^\infty\!
  d\lambda \ \ex^{iq^-\lambda} \ .
$$
This allows us to write
\begin{eqnarray}
  \int\! \frac{dq^+dq^-}{(2\pi)^2} \
  \  \delta(q^-)  \
  \frac{\ex^{- i (q^+ \x^- + q^- \xi^+)}}{( q^2+ i0 )[q^+]_{\rm ML}}
& = &
  \int\! \frac{dq^+dq^-}{(2\pi)^2} \
   \  \delta(q^-)  \
   \frac{\ex^{- i (q^+ \x^- + q^- \xi^+)}}{( q^2+ i0 )[q^+]_{\rm ML}}
\nonumber \\
\rule{0in}{4ex}
& = &
  \frac{-i}{(2\pi)^3} \ \int_0^\infty\! d\alpha
  \int_{-\infty}^\infty\! d\lambda\int\!\frac{dq^+dq^-}{[q^+]_{\rm ML}}
\nonumber \\
\rule{0in}{4ex}
& & \times
  \exp\[i(\alpha q^2 + (\xi^+ + \lambda) q^- +  \xi^- q^+ )\] \
   \ .
\end{eqnarray}
%Eq (20)

To proceed, we make use of the results obtained in \cite{CDL83,BR05}
(that can be directly derived by applying Cauchy's theorem)
\begin{equation}
 \int\!\frac{dq^+dq^-}{[q^+]_{\rm ML}} \
 \exp\[i(\alpha q^+q^- +  \beta^+ q^- +\beta^- q^+)\] \
=
 -\frac{2\pi}{\beta^+}
 \[ \exp\(-i \frac{ \beta^+ \beta^-}{\alpha} \) - 1  \]
\label{eq:MLmaster}
\end{equation}
%Eq (21)
and employ the representation
\begin{equation}
  \frac{1}{\lambda} \(1 - \ex^{-i \xi^- \lambda}\)
=
  i \xi^- \ \int_0^1\!d\tau \ \ex^{- i \tau \xi^- \lambda}
\end{equation}
%Eq (22)
to find
\begin{equation}
  \vecc A^\perp (\x)
=
  - i \frac{g}{2} \int\! \frac{d^2 q_\perp}{(2\pi)^2}
  \frac{\vecc q^\perp}{\vecc q_\perp^2} \,
  \ex^{i \mbox{\footnotesize\boldmath$ q_\perp$}
  \cdot
  \mbox{\footnotesize\boldmath$\xi_\perp$}
      } \
  \( \frac{\xi^-}{2\pi} \ \int_0^1\!d\tau \
  \int_{-\infty}^\infty\! d\lambda \ \ex^{- i \xi^- \lambda \tau} \)
  \ .
\end{equation}
%Eq (23)
Evaluation of the pure transverse integral yields
\begin{equation}
  \int\! \frac{d^2 q_\perp}{(2\pi)^2}
  \frac{\vecc q^\perp}{\vecc q_\perp^2} \,
  \ex^{i{\mbox{\footnotesize\boldmath$ q_\perp$}
  \cdot
  \mbox{\footnotesize\boldmath$\xi_\perp$}
      }}
=
  - \frac{i}{2\pi} \
  \vecc \nabla^\perp \ln \Lambda |\vecc  \x_\perp| \ ,
\label{eq:trans-expr}
\end{equation}
%Eq (24)
where $\Lambda$ is again an auxiliary IR regulator which shall
ultimately drop out from all physical quantities.
Therefore, the transverse gauge field at light-cone infinity in the
ML-gauge reads
\begin{equation}
   \vecc A^\perp (\infty^-; \vecc \x_\perp) = - \frac{g}{4\pi}
   \vecc \nabla^\perp \ln \ \Lambda |\vecc \x_\perp| \, .
\label{eq:MLret}
\end{equation}
%Eq (25)
Hence, imposing to the evaluation of the gluon propagator
the ML-prescription, the transverse gauge field at light-cone
infinity is given by Eq. (\ref{eq:MLret}) and is a total transverse
derivative, just as its counterpart (\ref{eq:ret}) in the case of
$q^-$-independent prescriptions.
But there is a crucial difference: in contrast to a $q^-$-independent
prescription, the ML result does not bear any dependence on the
imposed boundary conditions encoded in the constant $C_\infty$.
The resulting phase, accumulated by the struck quark moving along
the ``plus''-light-cone ray, can, therefore, be presented in a similar
way as in the case of the $q^-$-independent prescriptions
\cite{JY02,BJY03,BMP03}, viz.,
\begin{equation}
  \hbox{transverse phase}
=
  \pa \exp\[- i g \int_0^\infty\! d\t \ \vecc l^\perp
  \cdot \vecc A^\perp (\infty^-, 0^+; \vecc l \t )\] \ ,
\end{equation}
%Eq (26)
where $\vecc l$ is an arbitrary two-dimensional vector.
This result is crucial for our considerations in the next section.

\section{Calculation of the anomalous dimension in the
         ML-gauge}
\label{sec:anom-dim-ML}

\begin{figure}[h]
\centering
\includegraphics[scale=0.6,angle=0]{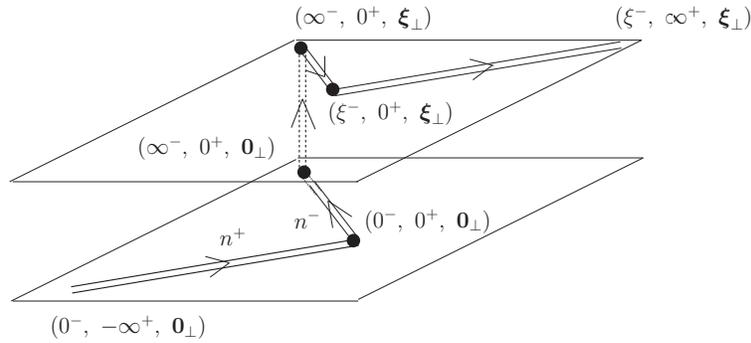}~~
\caption{The integration contour associated with the additional soft
         counter term.
\label{fig:contour}}
\end{figure}

\begin{figure}
\centering
\includegraphics[scale=0.7,angle=90]{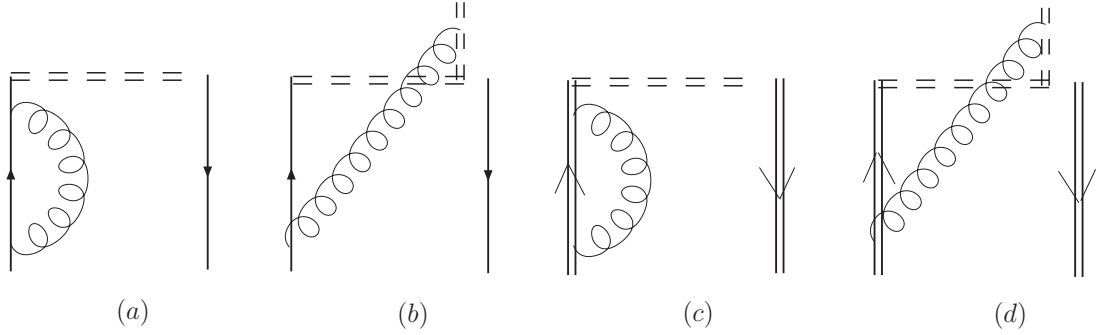}~~
\caption{Virtual one-loop gluon contributions (curly lines) to the
         UV-divergences of the TMD PDF in the light-cone gauge---graphs
         (a) and (b).
         The graphs (c) and (d) are corresponding contributions
         originating from the soft factor $R$.
         Double lines denote gauge links.
         The vertical ones represent the transverse gauge links.
         The Hermitian conjugated diagrams are not shown.
\label{fig:se_gluon}}
\end{figure}

\begin{figure}
\centering
\includegraphics[scale=0.7,angle=90]{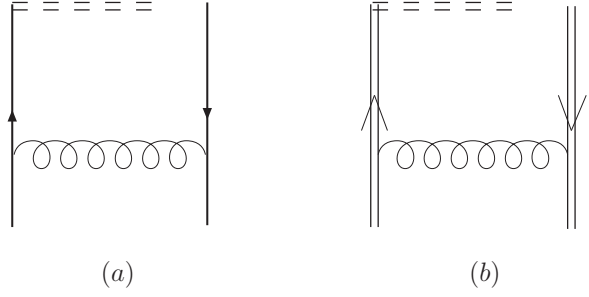}~~
\caption{Real gluon contributions (curly lines) to the TMD PDF in the
         light-cone gauge using the ML pole prescription
         (``ML gauge'').
         Double lines denote gauge links.
         The Hermitian conjugated diagrams are not shown.
\label{fig:se_real-gluon}}
\end{figure}

We start our anomalous-dimension considerations by recalling the
modified definition of the TMD PDF, proposed in Refs. \cite{CS07,CS08}:
\begin{eqnarray}
  f_{q/q}^{\rm mod}\left(x, \mbox{\boldmath$k_\perp$};\mu\right)
&& \!\!\! =
  \frac{1}{2}
  \int \frac{d\xi^- d^2\mbox{\boldmath$\xi_\perp$}}{2\pi (2\pi)^2}
  {\rm e}^{-ik^+\xi^- +i { \mbox{\footnotesize\boldmath$k_\perp$}}
  \cdot \mbox{\footnotesize\boldmath$\xi_\perp$}}
  \left\langle
              q(p) |\bar \psi (\xi^-, \mbox{\boldmath$\xi_\perp$})
              [\xi^-, \mbox{\boldmath$\xi_\perp$};
   \infty^-, \mbox{\boldmath$\xi_\perp$}]^\dagger
\right. \nonumber \\
\rule{0in}{3ex}
&& \left. \times
   [\infty^-, \mbox{\boldmath$\xi_\perp$};
   \infty^-, \mbox{\boldmath$\infty_\perp$}]^\dagger
   \gamma^+[\infty^-, \mbox{\boldmath$\infty_\perp$};
   \infty^-, \mbox{\boldmath$0_\perp$}]
   [\infty^-, \mbox{\boldmath$0_\perp$}; 0^-,\mbox{\boldmath$0_\perp$}]
\right. \nonumber \\
\rule{0in}{4ex}
&& \left. \times
   \psi (0^-,\mbox{\boldmath$0_\perp$}) |q(p)
   \right\rangle
   R (p^+, n^-|\xi^{-}, \mbox{\boldmath$\xi_\perp$}) \ \, .
\label{eq:tmd_re-definition}
\end{eqnarray}
%Eq (27)
This definition differs from the standard one, given by
Eq. (\ref{eq:tmd_naive}), because it takes into account an additional
soft factor $R$ which is defined as the vacuum expectation value of
the gauge links \cite{CS07,CS08}
\begin{equation}
  R
\equiv
 \left\langle 0
 \left| {\cal P} \exp\Big[ig \int_{\Gamma_{\rm cusp}}d\zeta^\mu
 \ t^a A^a_\mu (\zeta)\Big] \cdot
 {\cal P}^{-1} \exp\Big[-ig \int_{\Gamma_{\rm cusp}'}d\zeta^\mu
 \ t^a A^a_\mu (\xi + \zeta)\Big]
 \right|0
 \right\rangle\ ,
\label{eq:soft_factor_1}
\end{equation}
%Eq (28)
illustrated in Fig.\ \ref{fig:contour}, and with the involved contours
being defined by
\begin{eqnarray}
&& \nonumber \Gamma_{\rm cusp} : \ \zeta_\mu
=
  \{ [p_\mu^{+}s \ , \ - \infty < s < 0] \
 \cup \ [n_\mu^-  s' \ ,
  \ 0 < s' < \infty] \ \cup \
  [ \mbox{\boldmath$l_\perp$} \tau , \, \ 0 < \tau < \infty ] \}
\\
&& \Gamma_{\rm cusp}' : \ \zeta_\mu
=
  \{ [p_\mu^{+}s \ , \ + \infty < s < 0] \
 \cup \ [n_\mu^-  s' \ ,
  \ 0 < s' < \infty] \ \cup \
  [ \mbox{\boldmath$l_\perp$} \tau , \, \ 0 < \tau < \infty ] \}\ .
~~~~~
\label{eq:gpm}
\end{eqnarray}
%Eq (29)
%%%%%%%%%%%%%%%%%%%%%%%%%%%%%%%FIGURE 2%%%%%%%%%%%%%%%%%%%%%%%%%%%%%%%%
%%%%%%%%%%%%%%%%%%%%%%%%%%%%%%%%%%%%%%%%%%%%%%%%%%%%%%%%%%%%%%%%%%%%%%%
[Note that the result does not depend on the particular
choice of the vector $\mbox{\boldmath$l_\perp$}$.]
The introduction of the soft factor $R$ is necessitated by the
demand to cancel undesirable mixed rapidity divergences arising in
the calculations with light-like quantities
\cite{Col08,JMY04,CH00,Hau07,CM04,CS07,CS08}.
Indeed, we have shown in \cite{CS07,CS08}, using the light-cone gauge
with the advanced, retarded, or principal-value prescription, that the
anomalous dimension entailed by the UV-divergences of graph (c) in
Fig.\ \ref{fig:se_gluon} (generated by the soft factor $R$), cancels
the corresponding contribution of the TMD PDF, given by graph (a) in
the same figure.
On the other hand, pure rapidity divergences still appear in the
UV-finite graphs with real-gluon emissions---see Fig.\
\ref{fig:se_real-gluon}.

In this section, we shall show that the use of the ML-gauge allows
one to avoid rapidity divergences in the anomalous dimension of the TMD
PDF, while preserving at the same time the validity (and gauge
invariance) of definition (\ref{eq:tmd_re-definition}).
Nevertheless, the rapidity divergences, disentangled from the
UV-singularities, are still present in the ML-gauge as well.
The resummation of them should be pursued by means of the evolution
equation which is analogous to the Collins-Soper one.
This issue will be considered elsewhere separately.

Up to the LO in powers of $\alpha_s$, the TMD PDF
(\ref{eq:tmd_re-definition}) can be cast in the form
\begin{equation}
  f_{q/q}^{\rm LO}
=
  f^{(0)} + f^{(1)} + O(\alpha_s^2)\ \ , \ \
  f^{(1)} = f_{\rm virt.}^{(1)} + f_{\rm real}^{(1)} \ ,
\end{equation}
%Eq (30)
where we have separated virtual (see Fig.\ \ref{fig:se_gluon}) from
real (see Fig.\ \ref{fig:se_real-gluon}) corrections (labeled
accordingly).
The real-gluon terms $f_{\rm real}^{(1)}$ do not contain UV divergences
and hence will not be considered any further.
In the tree approximation, one has
\begin{eqnarray}
  f_{q/q}^{(0)} (x, \vecc k_\perp)
& = &
  \frac{1}{2}
  \int \frac{d\xi^- d^2\mbox{\boldmath$\xi_\perp$}}{2\pi (2\pi)^2}
  {\rm e}^{-ik^+\xi^- +i { \mbox{\footnotesize\boldmath$k_\perp$}}
  \cdot \mbox{\footnotesize\boldmath$\xi_\perp$}}
  \langle
              p |\bar \psi (\xi^-, \mbox{\boldmath$\xi_\perp$})
   \gamma^+
   \psi (0^-,\mbox{\boldmath$0_\perp$}) |p \rangle
\nonumber \\
& = &
  \delta(1-x) \delta^{(2)} (\vecc k_\perp) \ .
\end{eqnarray}
%Eq (31)

%%%%%%%%%%%%%%%%%%%%%%%%%%%%%%%FIGURE 3%%%%%%%%%%%%%%%%%%%%%%%%%%%%%%%%
%%%%%%%%%%%%%%%%%%%%%%%%%%%%%%%%%%%%%%%%%%%%%%%%%%%%%%%%%%%%%%%%%%%%%%%

The extraction of the UV-singular part of $f^{(1)}$ proceeds
along the lines of our previous works, described in \cite{CS07,CS08}.
In order to isolate the leading-order UV-divergent terms, one has to
consider the virtual one-gluon contributions depicted in the diagrams
of Fig.\ \ref{fig:se_gluon}.
These diagrams amount to

$$
  f_{\rm virt.}^{(1)}
=
   \delta(1-x)
   \delta^{(2)} (\vecc k_\perp) \ \Sigma^{(1)}_{\rm virt.} (p) \
   \gamma^+ \ ,
$$
so that
$\Sigma^{(1)}_{\rm virt.} = \Sigma^{(a)} + \Sigma^{(b)}$.
The quark self-energy diagram $(a)$ gives (in dimensional
regularization with $ \w = 4 - 2 \e $)
\begin{equation}
   {\Sigma}^{(a)} (p, \a_s ; \mu, \e)
=
   {\Sigma}^{(a)}_{\rm Feynman} + {\Sigma}^{(a)}_{\rm ML}
= -g^2 C_{\rm F} \m^{2\e}\ \int\! \frac{d^\w q}{(2\pi)^\w}
   \frac{\g_\m (\hat p - \hat q) \g_\n}{(p-q)^2
   (q^2  + i0)}\  d^{\m\n}_{\rm ML}(q)
   \frac{i\hat p}{p^2}
\label{eq:self-energ}
\end{equation}
%Eq (32)
with
\begin{equation}
  d^{\m\n}_{\rm LC} (q)
=
    g^{\m\n}
  - \frac{q^\m (n^{-})^\n + {q^\n (n^{-})^\m}}{[q^+]_{\rm ML}}\ .
\label{eq:gluon_pr}
\end{equation}
%Eq (33)
After some standard calculations, one has for the
prescription-independent $g_{\mu\nu}$ (``Feynman'') term
\begin{eqnarray}
  {\Sigma}^{(a)}_{\rm Feynman} (p, \a_s,  \mu , \e )
& = &
-g^2 C_{\rm F} \m^{2\e}\ \int\! \frac{d^\w q}{(2\pi)^\w}
   \frac{\g_\m (\hat p - \hat q) \gamma^\mu}{(p-q)^2
   (q^2  + i0)}\
   \frac{i\hat p}{p^2}
\nonumber \\
& = &
    - \frac{\a_s}{4\pi} C_{\rm F} \  \G(\e)
    \(-4\pi\frac{\mu^2}{p^2}\)^\e
    \frac{\Gamma(1-\e) \Gamma(2- \e)}{\Gamma(2 - 2\e)}
\ .
\label{eq:fey3}
\end{eqnarray}
%Eq (34)

%%%%%%%%%%%%%%%%%%%%%%%%%%%%%%%FIGURE 4%%%%%%%%%%%%%%%%%%%%%%%%%%%%%%%%
%%%%%%%%%%%%%%%%%%%%%%%%%%%%%%%%%%%%%%%%%%%%%%%%%%%%%%%%%%%%%%%%%%%%%%%

The evaluation of the ML-dependent part
\begin{equation}
   {\Sigma}^{(a)}_{\rm ML} (p, \a_s, \m  ; \e )
=
   g^2 C_{\rm F} \m^{2\e}\!
   \int\! \frac{d^\w q}{(2\pi)^\w}
   \frac{1}{(p-q)^2 (q^2 + i0 )}
   \left[
      \frac{\hat q (\hat p - \hat q) \gamma^+ }{[q^+]_{\rm ML}}
     +\frac{\gamma^+ (\hat p - \hat q) \hat q}{[q^+]_{\rm ML}}
   \right]
     \frac{i \hat p}{p^2}
\label{eq:pole1}
\end{equation}
%Eq (35)
is more involved.
After the transformation of the numerator, one gets
\begin{equation}
    {\S}^{(a)}_{\rm ML}
=
    g^2 C_{\rm F} \m^{2\e}\ \int\! \frac{d^\w q}{(2\pi)^\w}
    \[ ( \hat p \g_\m \gamma^+ + \gamma^+ \g_\m \hat p)
    \frac{(p - q)^\m}{(p-q)^2}
  -2\gamma^+ \] \frac{1}{(q^2 +i0 ) [q^+]_{\rm ML}}
    \frac{i \hat p}{p^2}\ .
\label{eq:pole2}
\end{equation}
%Eq (36)
Let us first consider the following integral \cite{BR05, CDL83}:
\begin{eqnarray}
 \int\!\frac{d^\omega q}{(2\pi)^\omega}
 \frac{1}{[q^+]_{\rm ML}} \
 \frac{1}{q^2 (p-q)^2} \
& = &
 \int\! d\alpha \int\! d\beta \
 \int\!\frac{d^\omega q}{(2\pi)^\omega}
       \frac{\ex^{i(\alpha q^2 + \beta (p-q)^2)}}{[q^+]_{\rm ML}}
\nonumber \\
& = &
  - \frac{i}{ (4\pi)^{\omega/2} }
  \frac{\Gamma(\e)}{p^+ (-p^2)^\e} \frac{1}{\e} \
  \[\frac{\Gamma^2(1- \e)}{\Gamma(1- 2\e)} - 1 \]
  \ ,
\label{eq:MLmaster_1}
\end{eqnarray}
%Eq (37)
where we have assumed that the direction of the momentum $p$ of the
struck quark is purely longitudinal, i.e., $\vecc p_\perp = 0$.
At first sight, the above expression seems to have a double pole in
$1/\e$.
Expanding the $\Gamma$ functions, one, however, finds that Eq.\
(\ref{eq:MLmaster_1}) is finite and does not contribute any UV
singularities.
In contrast, the integral with $q^\mu$ in the numerator is UV singular.
In order to calculate it, we use the $\alpha$-representation to obtain
\begin{equation}
  \int\!\frac{d^\omega q}{(2\pi)^\omega}
  \frac{q^\mu}{[q^+]_{\rm ML}} \
  \frac{1}{q^2 (p-q)^2} \
=
  - \frac{i}{2}\ \int_0^1\!
  \frac{dx}{x} \ \int_0^\infty\! dL \
  \ex^{i x L p^2} \ \frac{\pd}{\pd p_\mu}\int\!
  \frac{d^\omega q}{(2\pi)^\omega}
  \frac{\ex^{i( L  q^2 - 2 x L (p \cdot q) )}}{[q^+]_{\rm ML}}
  \ .
\label{eq:ML_3}
\end{equation}
%Eq (38)
Taking into account that
\begin{equation}
   \frac{\pd}{\pd p_\mu} \frac{1}{p^+} = \frac{(n^-)^\mu}{(p^+)^2} \ ,
\end{equation}
%Eq (39)
one finds that there are two parts: one proportional to $(n^-)^\mu$,
the other to $p^\mu$.
The first part vanishes in the final result by virtue of
$(n^-)^\mu (n^-)_\mu = 0$.
Evaluating the second part, using (\ref{eq:MLmaster_1}), gives
\begin{equation}
  \int\!\frac{d^\omega q}{(2\pi)^\omega}
  \frac{q^\mu}{[q^+]_{\rm ML}} \ \frac{1}{q^2 (p-q)^2} \
=
  - \frac{i}{  (4\pi)^{\omega/2} }
  \frac{p^\mu \ \Gamma(\e)}{p^+ (-p^2)^\e} \
  \frac{\Gamma^2(1- \e)}{\Gamma(2- 2\e)}
  \ .
\label{eq:ML_4}
\end{equation}
%Eq (40)
This integral contains a single $1/\e$-pole and thus contributes to the
leading UV-singularity.
In total, we find for diagram (a)
\begin{equation}
  {\Sigma}^{(a)} (p, \a_s ; \mu, \e)
=
  {\Sigma}^{(a)}_{\rm Feynman} + {\Sigma}^{(a)}_{\rm ML}
=
  - \frac{\a_s}{4\pi} C_{\rm F} \  \G(\e)
  \(-4\pi\frac{\mu^2}{p^2}\)^\e
  \frac{\Gamma^2(1-\e) }{\Gamma(2 - 2\e)} \[ (1-\e) - 4\] \ .
\label{eq:sigma-40}
\end{equation}
%Eq (41)
Extracting the UV divergent terms in the $\overline{\rm MS}$-scheme,
one gets (after adding the conjugated diagrams):
\begin{equation}
  {\S}_{(a)}^{\rm UV} (p, \a_s,  \m , \e )
=
  - \frac{\a_s}{4\pi} C_{\rm F} \[\frac{1}{\e}
    ( 1 - 4) - \gamma_E + 4\pi\]
=
  - \frac{3 \a_s}{4\pi} C_{\rm F} \[ \frac{1}{\e} - \gamma_E + 4\pi\]
   \ .
\label{eq:fey4}
\end{equation}
%Eq (42)
This expression makes it apparent that in the ML-gauge the UV-divergent
part of the TMD PDF (as well as the finite one) do not contain any
extra terms
of the form $\ln p^+$ which could be related to a cusped contour---in
contrast to the results obtained using
$q^-$-independent prescriptions \cite{CS07,CS08}.
Moreover, one sees that there is no imaginary term, as well, which is,
however, necessary in order to reproduce the result in covariant
gauges.
We shall show next how this term arises due to the transverse gauge
link at light-cone infinity.

\section{Contribution of the transverse gauge link at
         light-cone infinity}
\label{sec:trans-gauge-link}

The path-ordered composite transverse gauge link at light-cone infinity
reads
\begin{equation}
 \pa \exp\[+ i g \int_0^\infty\! d\t \vecc l^\perp
 \cdot \vecc A^\perp (\infty^-, 0^+;
 \vecc l_\perp \t + \vecc \x_\perp)\] \
 \pa \exp\[- i g \int_0^\infty\! d\t \vecc l^\perp
 \cdot \vecc A^\perp (\infty^-, 0^+; \vecc l_\perp \t )\] \ .
\label{eq:trans-gauge-link}
\end{equation}
%Eq (43)
In leading non-vanishing order, the corresponding diagram $(b)$ in
Fig.\ \ref{fig:se_gluon} yields
\begin{eqnarray}
   {\S}_{\rm ML}^{(b)} (p, \m , g; \e)
& = &
  - g^2 C_{\rm F} \m^{2\e} \int\!
   \frac{d^\w q'}{(2\pi)^\w} \!
   \int\! \frac{dq^+}{2\pi}
   \ex^{- i q^+ \infty^-} \!\!\int\! \frac{d^2 q_\perp}{(2\pi)^2} \
   \vecc l^\perp  \cdot \<0 | A^\m (q) \vecc A^\perp (q^{\prime})
                       |  0 \>
\nonumber \\
& & \times
   \frac{i}{(\vecc q^\perp \cdot \vecc l^\perp) + i0} \
   \frac{\gamma^+ (\hat p - \hat q) }{(p-q)^2}
 \ .
\label{eq:perp_LO}
\end{eqnarray}
%Eq (44)
To evaluate this expression, we employ the gluon propagator in the
ML-gauge which corresponds to the correlation function between the
longitudinal and the transverse gluon fields:
\begin{equation}
  \< 0 | A^\m (q) \vecc A^\perp (q^{\prime}) | 0 \>
=
  - \frac{\vecc q^\perp n^{-\m}}{ (q^2 + i0 ) [q^+]_{\rm ML}}
  (-i) (2\pi)^4 \d^{(4)} (q + q^{\prime}) \ .
\label{eq:trans_3}
\end{equation}
%Eq (45)
Using the explicit form of the ML-gauge field at light-cone infinity
(cf.\ Eq.\ (\ref{eq:MLret})),
the transverse integral can be rewritten in the form
\begin{equation}
  \int_0^\infty\! d\t \vecc l^\perp \cdot
  \vecc A^\perp (\infty^-, 0^+; \vecc l_\perp \t)
=
  \int\! \frac{dq^+}{2\pi}
  \ex^{- i q^+ \infty^-}\int\! \frac{d^2 q_\perp}{(2\pi)^2} \
  \vecc l^\perp \cdot \vecc A^\perp (q)
  \frac{i}{(\vecc q^\perp \cdot \vecc l^\perp) + i0} \ .
\label{eq:trans_2}
\end{equation}
%Eq (46)
Taking into account that
\begin{equation}
  \frac{1}{[q^+]_{\rm ML}}
=
  \frac{q^-}{q^+q^- + i0}
  =
  q^- \[ {\cal P} \frac{1}{q^+q^-} - i \pi \delta (q^+q^-) \] \ ,
\end{equation}
%Eq (47)
and using the equation (valid in the sense of distributions, see,
e.g., Ref. \cite{BJY03})
\begin{equation}
  \frac{\ex^{- i q^+ \infty^-}}{q^+ + i0}
  =
  - 2\pi i \ \delta (q^+) \ ,
\end{equation}
%Eq (48)
one can change variables in the $\delta$-function to obtain
\begin{eqnarray}
   {\S}_{\rm ML}^{(b)} (p, \m , g; \e)
=
   - g^2 C_{\rm F} \m^{2\e} 2\pi i  \int\!
   \frac{d^\w q}{(2\pi)^\w} \d (q^+) \
   \frac{\gamma^+ (\hat p - \hat q) }{(p-q)^2 \ q^2 }
 \ .
\label{eq:perp}
\end{eqnarray}
%Eq (49)
Finally, by taking the sum of the UV-divergent $(a)$ and $(b)$
contributions (Fig.\ \ref{fig:se_gluon}), we find
\begin{equation}
  {\S}^{(a+b)\rm UV}_{\rm ML} (p, \m, \a_s ; \e)
  =
  - \frac{\a_s}{\pi}\ C_{\rm F}\  \Bigg\{ \frac{1}{\e}
    \[\frac{1}{4} - \frac{ \gamma^+ \hat p}{2 p^+}
    \( 1   - \frac{i\pi}{2} \) \] - \gamma_E + 4 \pi \Bigg\}
\label{eq:s_tot}
\end{equation}
%Eq (50)
which yields
($
  \gamma^{+} \hat p \gamma^{+}/2 p^{+} = \gamma^{+}
$)
\begin{equation}
  {\S}^{(a+b)\rm UV}_{\rm ML} (p, \m, \a_s ; \e)
  =
  \frac{\a_s}{\pi}\ C_{\rm F}\  \[ \frac{1}{\e}
    \(\frac{3}{4}
    + \frac{i\pi}{2}  \) - \gamma_E + 4\pi \]
    \ .
\label{eq:s_tot-final}
\end{equation}
%Eq (51)
This result resembles what one finds in covariant gauges.
After including the mirror contribution to graph (b) in Fig.\
\ref{fig:se_gluon}, one obtains the following expression
\begin{equation}
  {\S}^{(a+b)\rm UV}_{\rm ML} (p, \m, \a_s ; \e)
  =
  \frac{\a_s}{\pi}\ C_{\rm F}\  \[ \frac{1}{\e}
    \frac{3}{4} - \gamma_E + 4\pi \]
    \ ,
\label{eq:s_tot-final}
\end{equation}
%Eq (52)
which is analogous to Eq.\ (\ref{eq:fey4}) and does not contain an
imaginary part.
Hence, for the Mandelstam-Leibbrandt pole prescription, the UV-singular
parts of the TMD PFDs reproduce the result obtained in a covariant
gauge, where there are no effects from artifacts of gauge-contour
obstructions (one encounters when using the light-cone gauge in
association with the advanced, retarded, or principal-value
prescription).

\section{Evaluation of the soft factor}
\label{sec:soft-factor}

To complete our arguments, we have now to verify whether the
modified definition (\ref{eq:tmd_re-definition}), proposed in
\cite{CS07,CS08} using a light-cone gauge in conjunction with
the advanced, retarded, or PV prescription, remains valid in the
ML-gauge as well.
The main ingredient of this definition is a soft factor which
was introduced in order to compensate the extra (mixed) UV divergence
and associated anomalous dimension originating from a cusped contour.
However, the latter are absent in the ML-gauge, as we have shown above.
Therefore, we have to demonstrate that the soft factor in this case
does not jeopardize Eq.\ (\ref{eq:tmd_re-definition}).
In leading order, the UV singularities of the soft factor are generated
by the self-energy of the light-like gauge link and the one-gluon
exchanges between the light-like and the transverse gauge link (see
diagrams $(c)$ and $(d)$ in Fig.\ \ref{fig:se_gluon}, respectively).
Thus, one has
\begin{equation}
  \F_{\rm soft}^{\rm LO}
=
  \F_{\rm soft}^{(0)} + \F_{\rm soft}^{(1)} + O(\alpha_s^2) \ \ ,
\label{eq:Phi-soft}
\end{equation}
%Eq (53)
$$
  \F_{\rm soft}^{(0)} = 1 \ \ ,
  \ \ \F_{\rm soft}^{(1)}
=
  \F_{\rm soft-virt}^{(1)} + \F_{\rm soft-real}^{(1)} \ , \
  \F_{\rm soft-virt}^{(1)}
=
  \F_{\rm soft-virt}^{(c)} + \F_{\rm soft-virt}^{(d)}
$$

\begin{eqnarray}
  \F_{\rm soft-virt}^{(c)}
& = &
  i g^2 \m^{2\e} C_{\rm F}\ u_\m u_\n
  \int_0^\infty d\s \int_0^\s d\t \int\! \frac{d^\w q}{(2\pi)^\w}
  \frac{\ex^{-i q \cdot u (\s-\t)}}{q^2 + i0}
  \(g^{\m\n} - \frac{q^\m n^{-\n} + q^\n n^{-\m}}{[q^+]} \) \
\nonumber \\
& = &
   2 i g^2 \m^{2\e} C_{\rm F}\
  \int_0^\infty d\s \int_0^\s d\t \int\! \frac{d^\w q}{(2\pi)^\w}
  \frac{\ex^{-i q^- (\s-\t)}}{2 q^+q^- - \vecc q_\perp^2 + i0}
   \frac{q^-}{q^+ + i0 q^-}  \ ,
\label{eq:soft-factor1}
\end{eqnarray}
%Eq (54)
where the vector $u_\mu$ is chosen to be light-like:
$u_\mu = (p^+, 0^-, \vecc 0_\perp)$.
Due to the relative positions of the poles in the Feynman and the
ML-denominators, this integral is zero \cite{BKKN93}, i.e.,
both poles are on the same side of the $q^+$-axis:
\begin{equation}
  \F_{\rm soft-virt}^{(c)\rm ML}
  =
  0 \ .
\end{equation}
%Eq (55)
For the same reason, the contribution of diagram $(d)$ in Fig.\ 2
vanishes as well, entailing $\F_{\rm soft-virt}^{(d)\rm ML} = 0$.
On the other hand, the contribution arising from real gluons,
$\F_{\rm soft-real}^{(1)}$, does not contain UV-singularities.
Hence, $\F_{\rm soft}^{\rm LO}$ reduces to unity, excluding the
appearance of any contribution to the anomalous dimension of the
TMD PDF related to spurious rapidity divergences.
This result validates Eq.\ (\ref{eq:tmd_re-definition}) also for
the case of the light-cone gauge with the ML-prescription.

Further, the anomalous dimension of the modified TMD PDF
(\ref{eq:tmd_re-definition}) coincides, therefore, with the anomalous
dimension of the standard TMD PDF (cf.\ (\ref{eq:tmd_naive}))
in the light-cone gauge with the ML-prescription.
Consequently,  the renormalization-group properties of the TMD PDF are
controlled by the following evolution equation
\begin{eqnarray}
  \frac{1}{2} \mu \frac{d}{d\mu}
  \  f_{i/q}^{\rm mod}(x, \mbox{\boldmath$k_\perp$}; \mu)
& & =
  \int\! d^2 \vecc q_\perp \ \int_x^1\! \frac{dz}{z} \ P_\perp
  \(\frac{x}{z}, \vecc q_\perp, \alpha_s \) \
  f_{i/q}^{\rm mod} (z, \vecc q_\perp, \mu) \ ,
  \nonumber \\
  P_\perp \(y, \vecc q_\perp, \alpha_s \)
& & =
  \gamma_{\rm ML} \ \delta(1-y) \
  \delta^{(2)} (\vecc k_\perp - \vecc q_\perp) + O(\alpha_s^2) \ ,
\end{eqnarray}
%Eq (56)
where the anomalous dimension coincides with the standard expression,
i.e.,
\begin{equation}
  \gamma_{\rm ML}
  =
  - \halb \mu \frac{d}{d\mu}
  \ln \Sigma_{\rm ML} (\alpha_s, \e)
=
  \frac{3}{4} \frac{\a_s}{\pi}\, C_{\rm F} + O(\a_s^2)\ .
\end{equation}
%Eq (57)

\section{Conclusions}
\label{sec:concl}

This work was devoted to the treatment of mixed rapidity divergences of
fully gauge-invariant TMD PDFs when employing the light-cone
gauge in conjunction with the Mandelstam-Leibbrandt pole prescription.
To this end, we calculated the leading-order contributions to the TMD
PDF ensuing from virtual gluon corrections.
Exactly these terms contain the UV singularities of the TMD PDF and
thereby entail its anomalous dimension.
We have shown by explicit calculation at one loop that, in contrast to
other popular pole prescriptions, like the advanced, retarded, or the
principal-value one, the Mandelstam-Leibbrandt prescription possesses
the important property that spurious mixed rapidity divergences, related
to obstructions of the gauge contour, are absent.
Correspondingly, the soft factor, we introduced in \cite{CS07,CS08}
in order to ensure the cancelation of such artifacts to the
anomalous dimension of the TMD PDF, reduces in this case to unity, thus
preserving its validity.

Phenomenologically, the use of the ML pole prescription in the 
light-cone gauge will facilitate calculations of TMD PDFs in a 
factorized description of SIDIS cross sections because the 
contributions to the anomalous dimensions from gauge-contour 
obstructions in the Wilson lines of the TMD PDFs cancel out, making the 
insertion of a correcting soft factor superfluous right from the start.
This is also reflected in the evolution behavior of the TMD PDFs which
is controlled by the standard anomalous dimension one finds in a
covariant gauge, where anomalous-dimension artifacts are manifestly
absent because factorization is complete like in collinear factorization.
These aspects and the practical analysis of their applications will be
considered in more detail elsewhere.

%%%%%%%%%%%%%%%%%%%%%%%%%%%%%%%%%%%%%%%%%%%%%%%%%%%%%%%%%%%%%%%%%%%%%%%
\acknowledgments
%%%%%%%%%%%%%%%%%%%%%%%%%%%%%%%%%%%%%%%%%%%%%%%%%%%%%%%%%%%%%%%%%%%%%%%
We would like to thank Oleg Teryaev for stimulating discussions and
useful remarks.
This investigation was partially supported by the Heisenberg-Landau
Programme (Grant 2009), the Deutsche Forschungsgemeinschaft under
contract 436RUS113/881/0, the Alexander von Humboldt-Stiftung, the RF
Scientific Schools grant 195.2008.9, and the INFN.
\bibliographystyle{apsrev}
%\bibliographystyle{prsty-ab}
%\bibliographystyle{apsrev-ab}

%%%%%%%%%%%%%%%%%%%%%%%%%%%%%%%%%%%%%%%%%%%%%%%%%%%%%%%%%%%%%%%%%%%%%%%

\end{document}